# What we *understand* is what we get: Assessment in Spreadsheets

Andrea Kohlhase
DFKI (German Center for Artificial Intelligence)
Enrique-Schmidt-Str. 5
28359 Bremen, Germany
andrea.kohlhase@dfki.de

Michael Kohlhase
Jacobs University Bremen
Campus Ring 1
28759 Bremen, Germany
m.kohlhase@jacobs-university.de

**ABSTRACT**

*In previous work we have studied how an explicit representation of background knowledge associated with a specific spreadsheet can be exploited to alleviate usability problems with spreadsheet-based applications. We have implemented this approach in the SACHS system to provide a semantic help system for spreadsheets applications. In this paper, we evaluate the (comprehension) coverage of SACHS on an Excel-based financial controlling system via a "Wizard-of-Oz" experiment. This shows that SACHS adds significant value, but systematically misses important classes of explanations. For judgements about the information contained in spreadsheets, we provide a first approach for an "assessment module" in SACHS.*

**1 INTRODUCTION**

Spreadsheets are great active documents, they are intuitive, flexible, and offer a direct approach to computation. Unfortunately, an obverse statement is equally true, as they are error-prone but high-impact, widely-disseminated but poorly documented, and contain actual data in legacy form (see e.g. [Panko, 2000],[Murphy, 2008]). Support for comprehending spreadsheets is often concerned with data visualization techniques and data/formula dependency graphs (see [Brath & Peters, 2008] and [Hodnigg & Mittermeir, 2008] as examples). User assistance (e.g. help systems) are valuable but still largely missing except for a documentation-through-annotation approach in [Dinmore, 2009] and conceptual recognition of an interpretation issue in [Banks & Monday, 2008].

In our previous research we addressed this issue with semantic technology resulting in the SACHS system [Kohlhase & Kohlhase, 2009a/b/c/d]. It is a semantic help system for "**DCS**", a financial controlling system based on Excel in daily use at the German Research Center for Artificial Intelligence (DFKI). Here, a spreadsheet is illustrated with a semi-formal ontology of the relevant background knowledge via an interpretation mapping. An **ontology** defines the terms used to describe and represent a certain knowledge area. Concretely, it contains knowledge in a structured form, particularly concepts and their relationships. The formal parts of the ontology are then used to control the aggregation of help texts (from the informal part of the ontology) about the objects in the spreadsheet.

HODNIGG and MITTERMEIR state that "*comprehension of a workbook is non-trivial as there are several factors that aggravate its comprehension.*" [Hodnigg & Mittermeir, 2008, p. 82]. But what are the necessary factors for comprehension? With the SACHS system in a usable state, we have evaluated coverage with a **"Wizard-of-Oz" experiment** — a research method in which subjects interact with a computer system that they believe to be autonomous, but which is actually being operated or partially operated by an unseen human being (see [Wikipedia, 2009]). Interestingly, the experiment has revealed that the DCS system only models the factual part of the situation it addresses, while important aspects for 'understanding the numbers' remain implicit — and as a consequence the SACHS system also fails to tackle them.

For instance, users often ask questions like "*Is it good or bad if this cell has value 0.992?*" and experienced controllers may tell users "*Cell E6 must always be higher than E15*". We consider this knowledge (which we call **assessment knowledge**) to be an essential part of the background knowledge to be modeled in the semantically enhanced spreadsheet systems, since we can only profit from help if it is understood in 'all' its consequences. In particular, the assessment knowledge must be part of user assistance (e.g. answering the first question) and can be used to issue warnings (e.g. if the controller's invariant inherent in the second statement is violated).

|    | A | B | C | D | E | F | G | H |
|----|---|---|---|---|---|---|---|---|
| 1  | Profit and Loss Statement | | | | | | | |
| 2  |   |   |   |   |   |   |   |   |
| 3  | (in Millions) | | | Actual | | | Projected | |
| 4  |   | 1984 | 1985 | 1986 | 1987 | 1988 | 1989 | 1990 |
| 5  |   |   |   |   |   |   |   |   |
| 6  | Revenues | 3,865 | 4,992 | 5,803 | 5,441 | 4,124 | 4,617 | 5,223 |
| 7  |   |   |   |   |   |   |   |   |
| 8  | Expenses | | | | | | | |
| 9  | Salaries | 0,285 | 0,337 | 0,506 | 0,617 | 0,705 | 0,805 | 0,919 |
| 10 | Utilities | 0,178 | 0,303 | 0,384 | 0,419 | 0,551 | 0,724 | 0,951 |
| 11 | Materials | 1,004 | 1,782 | 2,046 | 2,273 | 2,119 | 1,975 | 1,84 |
| 12 | Administration | 0,281 | 0,288 | 0,315 | 0,368 | 0,415 | 0,468 | 0,527 |
| 13 | Other | 0,455 | 0,541 | 0,674 | 0,772 | 0,783 | 0,794 | 0,805 |
| 14 |   |   |   |   |   |   |   |   |
| 15 | Total Expenses | 2,203 | 3,251 | 3,925 | 4,449 | 4,573 | 4,766 | 5,042 |
| 16 |   |   |   |   |   |   |   |   |
| 17 | Profit (Loss) | 1,662 | 1,741 | 1,878 | 0,992 | -0,449 | -0,149 | 0,181 |

Figure 1: A Simple (Extended) Spreadsheet after [Winograd, 1996 (2006)]

To keep the paper self-contained we give a short overview of the SACHS system in the next section, followed by a report about the "Wizard-of-Oz" experiment in Section 3. We will envision exemplarily how assessment knowledge can be used in the SACHS system in Section 4. Section 5 concludes the paper and discusses future research directions.

## 2 SACHS: A SEMANTIC HELP SYSTEM FOR MS EXCEL SPREADSHEETS

Even though Excel spreadsheets serve well as an interface for a financial controlling system, they are often too complex in practice. Thus, user assistance for high-impact spreadsheets makes sense to reduce this complexity. Concretely, we created a semantic help system called SACHS for DCS (our institute's financial controlling system in Excel form).

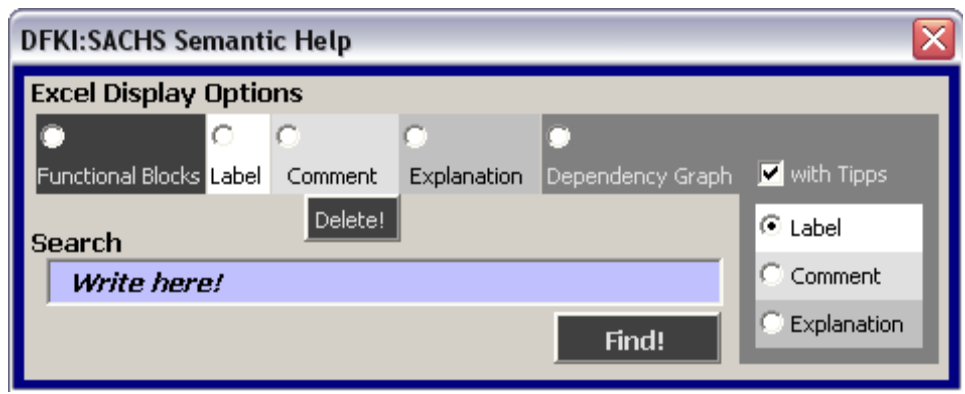

Figure 2: The SACHS Panel

In [Kohlhase & Kohlhase, 2009a] we analyzed spreadsheets as semantic documents. As such we diagnosed two meaningful layers — the surface structure and the formulae — with a strong Excel bias towards the latter. To compensate for this computational focus we proposed to augment the two existing semantic layers of a spreadsheet by one that makes the intention of the spreadsheet author explicit. Therefore we formalized implicit knowledge about the document into a background ontology.

Since the spreadsheet objects that carry meaning are the cells, we designed the interaction of the SACHS system as cell-based. Previously, cells were interpreted by the user via the grid layout (like within a table with an assigned row and column specification) and the underlying formula only. With SACHS we offered a third interpretation by aligning cells with concepts in the according background ontology. Hence, cell clicks are used as entry points for the help system: every click on a cell potentially generates help.

The SACHS panel shown in Fig. 2 offers the choice of getting either "**functional blocks**" (groups of cells belonging together), "**labels**" (titles), "**comments**" (short descriptions), or "**explanations**" (detailed descriptions). Note that our notion of "functional block" seems to be somewhat similar to ABRAHAM's "*regions*" [Abraham, 2005]. Generated help texts are enhanced by listing concrete cell values of dependent cells. For example, let us look at the simple spreadsheet in Fig. 1. The value in cell [*H*9] obviously depends on the respective year, here 1990. The generated SACHS label in Fig. 3 for cell [*H*9] thus contains a reference to the year 1990. Other optional help texts like comments and explanations are showcased in Fig. 3 as well.

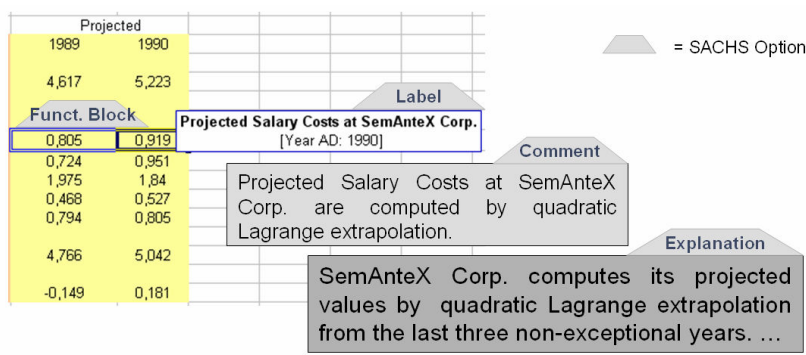

Figure 3: SACHS Help Display Options in Cell [H9]

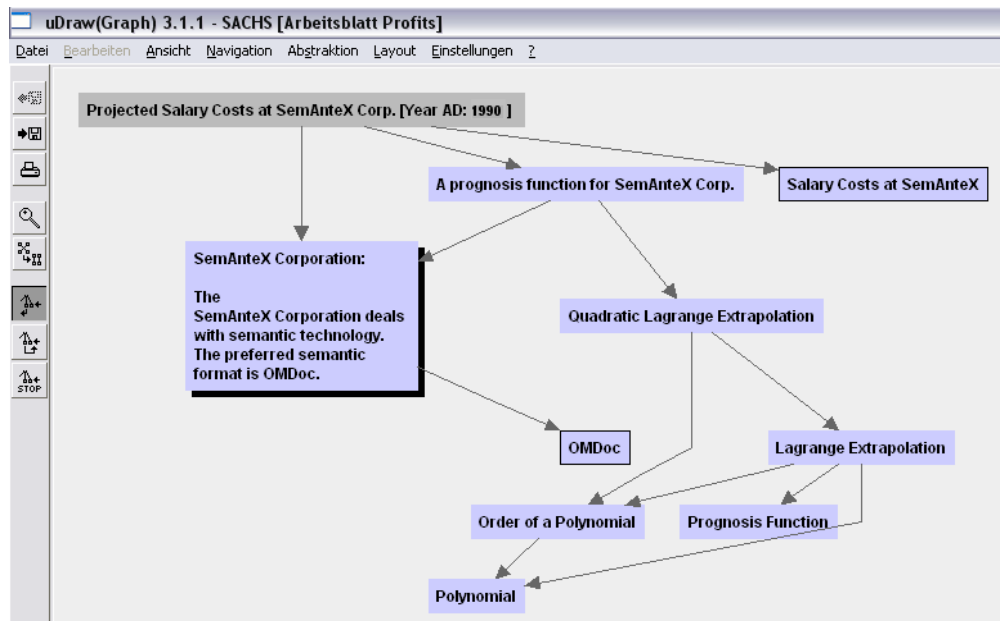

Figure 4: Dependency Graph Enabling Semantic Navigation

Another option in the SACHS panel is the generation of a **dependency graph** for the concept connected to the selected cell. For instance, if this option is chosen for cell [*H*9] (projected salary costs 1990), then the first two levels of the graph as seen in Fig. 4 are generated. If the user wants to elaborate on a specific concept, then a click on the corresponding node expands it by another level. This feature is comparable to hyperlinks in help texts, but adds **semantic navigation** cues (see [Kohlhase & Kohlhase, 2009d]). We mashed-up the graph-based interface with the interactions needed within a spreadsheet to allow the user to navigate the spreadsheet via the structured background ontology by the definitional structure of the intended functions. In the dependency graph the user has the choice of text granularity in each node (via right mouse click) or all nodes (via SACHS panel).

Another interesting extension of SACHS concerns "**framing**" (see [Kohlhase & Kohlhase, 2009d]): Does the user in Fig. 4 understand projected salary costs as some prognosis function or more specifically as quadratic Lagrange extrapolation? Generated help texts should vary according to these frames. Moreover, if the user frames cell [*H*9] as quadratic Lagrange extrapolation of first order, she might be interested in the computed cell value if the extrapolation were done with the Lagrange function of second order. Here, SACHS experiments with offering "**variants**".

## 3 HELP NEEDED, BUT WHERE?

To develop the domain ontology for the background knowledge of the DFKI controlling system DCS we organized interviews with a DFKI expert on the topic and recorded them as MP3 streams. We recorded three interview sessions amounting to approximately 1.5 hrs concerning 39 distinct knowledge items and containing 110 explanations. Even though these interviews were not originally intended as a "Wizard-of-Oz" experiment, we can use it as such with the DFKI expert in the role of the wizard and the interviewer 'interacting' with the expert's knowledge about the controlling system. In other words, the interviewee plays the part of an ideal SACHS system and gives help to the interviewer who plays the part of the user. This experiment gives us valuable insights about the *different qualities of knowledge* in a user assistance system, which the expert thought were necessary to understand the specific controlling system spreadsheet.

When studying the MP3 streams, we were surprised that in many cases a question of "*What is the meaning of ...*" was answered by the expert with up to six of the following **explanation types**, the occurrence rate of which relative to the number of knowledge items is listed in the brackets:

1. **Definition (Conceptual)** [71.8%]

A *definition* of a knowledge item like a functional block is a thorough description of its meaning. For example the functional block "cover ratio per project in a research area" was defined as the percentage rate to which the necessary costs are covered by the funding source and own resources.

2. **Purpose (Conceptual)** [46.2%]

The *purpose* of a knowledge item in a spreadsheet is defined by the spreadsheet author's intention, in particular, the purpose explains why the author put the information in. A principal investigator of a project or the respective department head e.g. needs to get the information about its cover ratio in order to know whether either more costs have to be produced to exploit the full funding money or more equity capital has to be acquired.

3. **Assessment of Purpose** [30.8%]

Given a purpose of a knowledge item in a spreadsheet, its reader must also be able to reason about the purpose, i.e., the reader must be enabled to draw the intended conclusions/actions or to *assess the purpose*. For understanding whether the cover ratio is as it is because not enough costs have yet been produced, the real costs have to be compared with the necessary costs. If they are still lower, then the costs should be augmented, whereas if they are already exploited, then new money to cover the real costs is needed.

4. **Assessment of Value** [51.3%]

Concrete values given in a spreadsheet have to be interpreted by the reader as well in order to make a judgement of the data itself, where this *assessment of the value* is a trigger for putting the assessment of purpose to work. For instance, the size of the cover ratio number itself tells the informed reader whether the project is successful from a financial standpoint. If the cover is close to 100%, "everything is fine" would be one natural assessment of its value.

5. **Formula** [23.1%]

With a given formula for a value in a spreadsheet's cell the reader knows exactly how the value was computed, so that she can verify her understanding of its intention against the author's. Note that a lot of errors in spreadsheets result from this distinction. In our experiment, if a value of a cell was calculated with a formula explicitly given in the spreadsheet, then the expert explained the dependency of the items in the formula, but restricted from just reading the formula aloud. In particular, he pointed to the respective cells and tried to convey the notion of the formula by visualizing their dependency, not so much what the dependency was about.

6. **Provenance** [43.6%]

The *provenance* of data in a cell describes how the value of this data point was obtained, e.g. by direct measurement, by computation from other values via a spreadsheet formula, or by import from another source; see [Moreau *et al.*, 2008] for a general discussion of provenance. In our interviews — as many of the data of the concrete spreadsheet were simply an output of the underlying controlling data base — the provenance explanations mostly referred to the specific data base where the data comes from. But when the formula for a value was computed, but not within Excel, the expert tried to give the formula as provenance information, e.g. in the case of the cover ratio. This knowledge was often very difficult to retrieve afterwards for the creation of the semantic document.

7. **History** [15.4%]

The *history*, i.e., the creation process of a spreadsheet over time, often is important to understand its layout that might be inconsistent with its intention. For instance, if an organizational change occurs that alleviates the controlling process and makes certain information fragments superfluous, then those fragments will still be shown in the transition phase and beyond, even though their entropy is now 100% in the most of cases.

These seven explanation types were distilled from the recorded set of 110 explanations. The percentages given can function as a *relevance ranking* done by the expert with respect to the importance of explanation types for providing help.

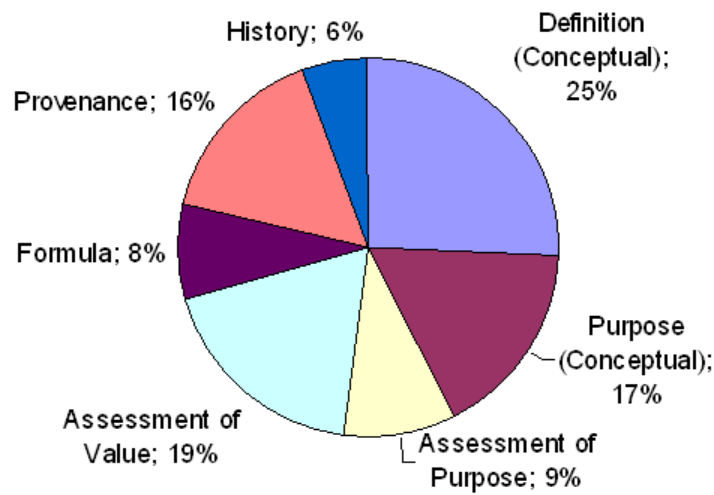

Figure 5: Help Needed — But Where?

Figure 5 portrays the distribution of occurrences according to each type. The "Wizard-of-Oz" experiment interpretation suggests that Fig. 5 showcases the user requirements for SACHS as a user assistance system (see also [Novick & Ward, 2006]). In particular, we can now *evaluate the SACHS system* with respect to this figure. Unsurprisingly, Definition explanations were the most frequent ones. Indeed, the SACHS system addresses this explanation type either with the dependency graph-based explanation interface in Fig. 4 or the direct help text generator shown in Fig. 3. But the next two types are not covered in the SACHS system, even though it can be argued that the ontology-based SACHS architecture is well-suited to cope with Purpose explanations — indeed, some of the purpose-level explanations have erroneously found their way into SACHS definitions, where they rather should have been classified as 'axioms and theorems' (which are currently not supported by the SACHS interface). The next explanation category (Provenance; 16%) has been anticipated in the SACHS architecture (see [Kohlhase & Kohlhase, 2009a]) but remains unimplemented in the SACHS system. The Assessment of Purpose type is completely missing from SACHS as well as Assessment of Value. Explanations of type Formula are only rudimentarily covered in SACHS by virtue of being a plugin that inherits the formula bar from its host application Excel, which has some formula explanation functionality. Finally, the explanation type History is also not yet covered in SACHS.

To summarize the situation: Excel is able to give help for 8% of the explanations we found in the help of a human expert. The implemented SACHS system bumps this up to 33%, while the specified SACHS system can account for 50%. Even though this is certainly an improvement, it leaves much more to be desired than we anticipated. In particular, we draw the conclusion that background knowledge that 'only' contains a domain ontology is simply not enough.

We will try to remedy parts of this in the remainder of this paper. In particular, we take up the problem of Assessment of Value explanations. On the one hand, it is ranked second in the list of explanation types with a stunningly high percentage of 51.3%, which can be interpreted as the second-best type of explanations from the point of view of our expert. On the other hand, the very nice thing about assessment for computational data is that we can hope for a formalization of its assessment in the form of formulas, which can be evaluated by e.g. Excel in turn.

## 4 MODELING ASSESSMENT

A first-hand approach of complementing spreadsheets with assessment knowledge could be the inclusion of Assessment of Value information into the definition text itself. In the concrete SACHS ontology we felt that we had no other choice in order to convey as much knowledge as possible, it is ontologically speaking a very impure approach (hence wrong) as such judgements do not solely depend on the concept itself. For instance, they also depend on the respective Community of Practice: At one institution e.g. a cover ratio of 95% might be judged as necessary, at another 100% (or more) might be expected.

Therefore, first let us have take a closer look at assessment itself: What is it about? Assessments consist of value judgements passed on situations modeled by (parts of) spreadsheets. As such, we claim that assessments are deeply in the semantic realm. To strengthen our intuition, let us consider some examples; we will use a slightly varied version of the simple spreadsheet document in Fig. 1, which we have already used in [Kohlhase & Kohlhase, 2009a/d] for this. The following can be considered typical assessment statements:

1. "*Row [6] looks good.*"
2. "*The revenues look good.*"
3. "*I like this* [points to cell [E17]] *but that* [points to cell [F17]] *is a disaster.*"
4. "*I like the profit in 1987 but of course not that in 1988.*"
5. "*Upper Management will be happy about the leftover funds in [nn] that they can now use elsewhere, but the PI of the project will be angry that he got less work out of the project than expected. Not to mention the funding agency; they cannot be told of this at all, because it violates their subsistence policy.*"

On the surface, the first statement refers to a row in the spreadsheet, but if we look more closely, we see that this cannot really be the case, since if we shift the whole spreadsheet by one row, then we have to readjust the assessment. So it has to be about the intended meaning of row [6], i.e., the development of revenues over the years. Indeed we can paraphrase I) with II) — another clue that the assessments are really about situations modelled by a functional block in the spreadsheet. But assessments are not restricted to functional blocks as statements III) and IV) only refer to individual cells. Note again that the statements are not about the numbers 0.992 and -0.449 (numbers in themselves are not good or bad, they just are). Here, the assessment seems to be intentional, i.e., about the intention "the profit in 1987/8" rather than the extension.

Another way to view this is that the latter two assessments are about the argument/value pairs ⟨1987, 0.9920⟩ and ⟨1988, −0.4490⟩. We will make this view the basis of our treatment of assessment in SACHS: We extend the background ontology by statements that judge the intended functions in the functional blocks of the spreadsheet on their functional properties. The theoretical work is carried out elsewhere, we only want to demonstrate the usefulness of taking the results of the "Wizard-of-Oz" experiment seriously. Therefore, we now envision a SACHS extension dealing with assessment as part of our user assistance system based on these ideas.

## 4.1 THE ENVISIONED ASSESSMENT EXTENSION IN SACHS

We will now show how assessments can be made useful for the user. The assessments are bound to (the intended function of) a functional block, so we can extend the context menu with entries for available assessment functions. In Fig. 6 we assume a right mouse click on the cell [B17] to show the context menu with two assessment functions — one for assessing its absolute value ("Assess Values of fBlock")and the other for its relative value ("Assess Domain of fBlock"). For example, we know that profit values have to be positive to be considered good, but also that a drop in profit over the covered domain of years (even if still positive) has to be evaluated as not good. So what happens when an assessment function is selected by a user? Then SACHS is put into a special 'assessment mode', which brings assessment information to the user's attention.

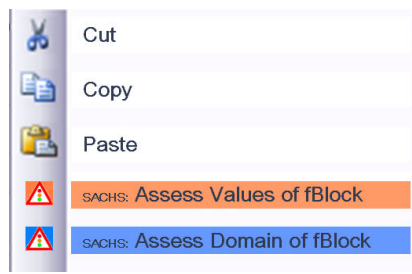

Figure 6: The Extended SACHS Context Menu

For instance, in Fig. 7 the user activated the absolute value assessment function. All cell values in the functional block of [B17] are positive except the one for the year 1988. SACHS color-codes this assessment to warn the user of any cells that get a negative judgement. At the same time, the assessment mode extends the explanatory labels by explanations texts from the assessment ontology.

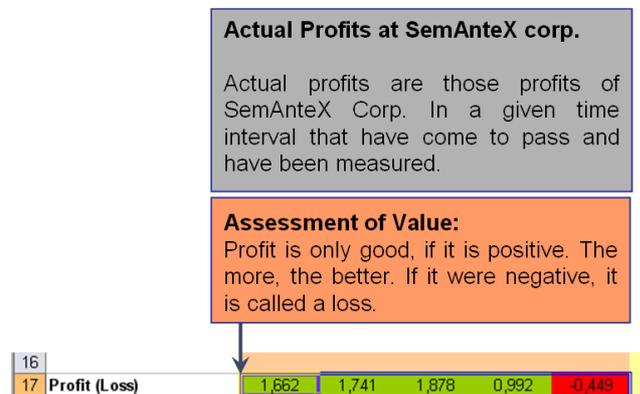

Figure 7: Assess the *Values*

A different color-coding is produced by SACHS, when the user activates the relative value assessment function (Fig. 8). Here, we realize that the profit has risen over the first considered three years, but started dropping in 1987. Thus, the first three cells are painted in green, whereas the last two of the functional block are painted in red.

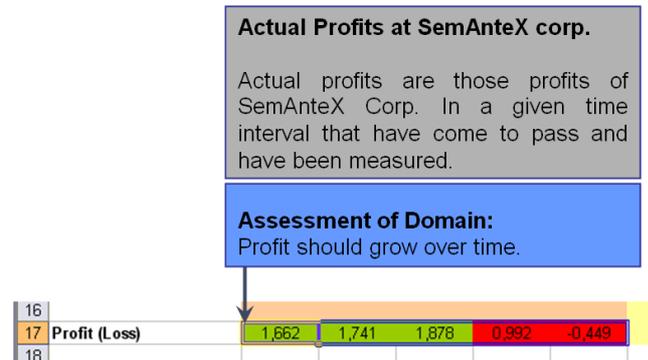

Figure 8: Assess the *Domain*

But as the assessments are synchronized with the assessment statements in the background ontology, in the dependency graph the user can analyze the assessments for possible causes. For example, recall that profits are defined as the difference between revenues and expenses. Then it makes sense to trace assessments through the dependency graph provided by the SACHS system. Note that this analysis is anchored to the cell:

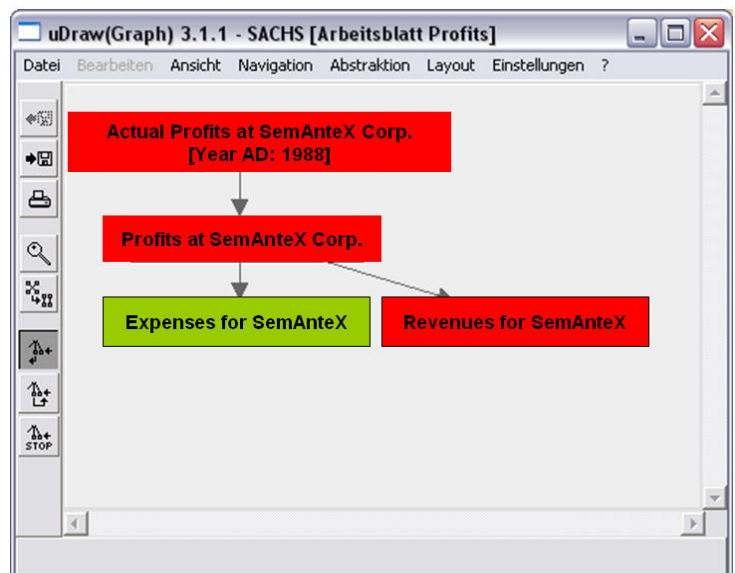

Figure 9: Assess *All* Values

Figure 9 shows the dependency graph for the negatively assessed cell [F17] for the profits in the year 1988.

Here the revenues are also negatively assessed (color-coded red in the graph), so the problem might be with the revenues. Note as well that this graph cannot be used for a causal analysis, as the arrows here define only dependency relations. We conjecture that causal analysis knowledge can transparently be included in the background ontology and can be made effective for the user in a similar interface.

## 5 CONCLUSION AND FURTHER WORK

In this paper we have reported an evaluation of SACHS, a semantic help system for a financial controlling system, via a (post-facto) "Wizard-of-Oz" experiment. The immediate results of this are twofold. The experiment basically validates the semantic approach implemented in the SACHS system: The availability of explicitly represented background knowledge resulted in a dramatic increase of the explanations that could be delivered by the help system. But the experiment also revealed that significant categories of explanations are still systematically missing from the current setup, severely limiting the usefulness of the system. We have shown how the SACHS system can be extended organically to include assessment functionalities, if the background ontology includes a formal model of assessment and conclude that the SACHS approach is sufficiently flexible to cover novel aspects of understanding spreadsheets.

An avenue for further research is the fact, that we have not yet made full use of the data from the "Wizard-of-Oz" experiment in section 3. For example, we could analyze the co-occurrences of distinct explanation types as seen in Fig. 10. We can ask *"Given an explanation of a certain type, then which other explanation type is needed or useful"?* A first dig into that research direction already yielded interesting results like an unusually high co-occurrence between Definition and Assessment of Purpose explanations. We plan to study these relationships further;

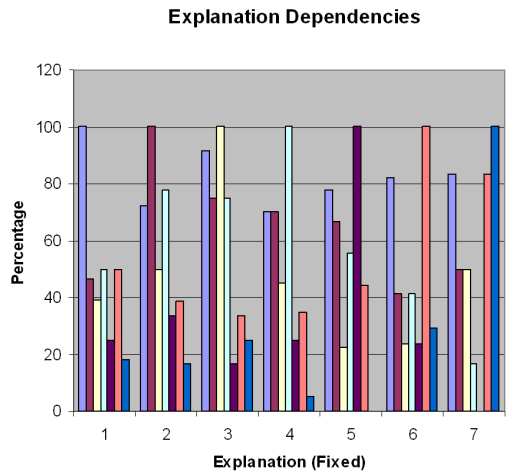

Figure 10: Explanation Dependencies

if these can be corroborated in other studies and other spreadsheet-based applications, we will fine-tune our text aggregation algorithm for the dependency graph interface in figure 4 to volunteer the experimentally correlated explanation types.

Finally, we observe that the conclusions of our "Wizard-of-Oz" experiment are neither restricted to the system Excel nor to the financial controlling domain.

# References

[Abraham, 2005] ABRAHAM, ROBIN. 2005. Template parsing with user feedback. *Pages 323–324 of: Vl/hcc.* IEEE Computer Society.
[Banks & Monday, 2008] BANKS, DAVID A., & MONDAY, ANN. 2008. Interpretation as a factor in understanding flawed spreadsheets, Proceedings of the EuSpRIG 2008 Symposium.
[Brath & Peters, 2008] BRATH, RICHARD, & PETERS, MICHAEL. 2008. Spreadsheet validation and analysis through content visualization, Proceedings of the EuSpRIG 2008 Symposium.
[Carette *et al.* , 2009] CARETTE, JACQUES, DIXON, LUCAS, SACERDOTI COEN, CLAUDIO, & WATT, STEPHEN M. (eds). 2009. *MKM/Calculemus 2009 proceedings*. LNAI, no. 5625. Springer Verlag.
[Dinmore, 2009] DINMORE, MATTHEW. 2009. Documenting problem-solving knowledge: Proposed annotation design guidelines and their application to spreadsheet tools, Proceedings of the EuSpRIG 2009 Symposium.
[Hodnigg & Mittermeir, 2008] HODNIGG, KARIN, & MITTERMEIR, ROLAND T. 2008. Metrics-based spreadsheet visualization: Support for focused maintenance, Proceedings of the EuSpRIG 2009 Symposium.
[Kohlhase & Kohlhase, 2009a] KOHLHASE, ANDREA, & KOHLHASE, MICHAEL. 2009a. Compensating the computational bias of spreadsheets with MKM techniques. *In:* [Carette *et al.* , 2009].
[Kohlhase & Kohlhase, 2009b] KOHLHASE, ANDREA, & KOHLHASE, MICHAEL. 2009b. Modeling task experience in user assistance systems. *In:* [Mehlenbacher *et al.* , 2009].
[Kohlhase & Kohlhase, 2009c] KOHLHASE, ANDREA, & KOHLHASE, MICHAEL. 2009c. Semantic transparency in user assistance systems. *In:* [Mehlenbacher *et al.* , 2009].
[Kohlhase & Kohlhase, 2009d] KOHLHASE, ANDREA, & KOHLHASE, MICHAEL. 2009d. Spreadsheet interaction with frames: Exploring a mathematical practice. *In:* [Carette *et al.* , 2009].
[Mehlenbacher *et al.* , 2009] MEHLENBACHER, BRAD, PROTOPSALTIS, ARISTIDIS, WILLIAMS, ASHLEY, & SLATTEREY, SHAUN (eds). 2009. *Proceedings of the 27$^{th}$ annual ACM international conference on design of communication,* ACM Press, for ACM Special Interest Group for Design of Communication.

[Moreau *et al.*, 2008]  MOREAU, LUC, GROTH, PAUL, MILES, SIMON, VAZQUEZ, JAVIER, IBBOTSON, JOHN, JIANG, SHENG, MUNROE, STEVE, RANA, OMER, SCHREIBER, ANDREAS, TAN, VICTOR, & VARGA, LASZLO. 2008. The provenance of electronic data. *Communications of the ACM*, **51**(4), 52–58.

[Murphy, 2008]  MURPHY, SIMON. 2008. Spreadsheet hell.

[Novick & Ward, 2006]  NOVICK, DAVID G., & WARD, KAREN. 2006. What users say they want in documentation. *Pages 84–91 of: SIGDoC'06 Conference proceedings*. ACM.

[Panko, 2000]  PANKO, RAYMOND R. 2000. Spreadsheet errors: What we know. what we think we can do. *In: Symp. of the European spreadsheet risks interest group (EuSRiG)*.

[Wikipedia, 2009]  WIKIPEDIA. 2009. *Wizard of Oz experiment — Wikipedia, the free encyclopedia*. [Online; accessed 20-May-2009].

[Winograd, 1996 (2006)]  WINOGRAD, TERRY. 1996 (2006). The spreadsheet. *Pages 228–231 of:* WINOGRAD, TERRY, BENNETT, JOHN, DE YOUNG, LAURA, & HARTFIELD, BRADLEY (eds), *Bringing design to software*. Addison-Wesley.